\documentclass[11pt]{article}
\usepackage{fancyhdr}
\usepackage[utf8]{inputenc}
\usepackage{graphicx}
\usepackage[super,comma,sort&compress]{natbib}
\usepackage{amssymb}
\usepackage{xcolor}
\usepackage{geometry}
\usepackage{amssymb} 
\usepackage{amsmath}
\definecolor{darkgreen}{rgb}{0,0.5,0}

\newif\ifusesup
\usesuptrue

\newcommand{\thetitle}{Cavity-mediated coupling of mechanical oscillators limited by quantum backaction}

\newcommand{\dcaption}[1]{\caption{#1}}

\begin{document}

\baselineskip 18pt

\title{\textbf{\thetitle}}
\author{
    \textsf{Nicolas Spethmann$^{1,2}$\thanks{Email: spethmann@berkeley.edu}, Jonathan\ Kohler$^{1}$, Sydney Schreppler$^1$, Lukas Buchmann$^1$,}\\\textsf{Dan M.\ Stamper-Kurn$^{1,3}$}
    \\ \\
    \small{1. Department of Physics, University of California, Berkeley, CA 94720, USA}\\
    \small{2. Fachbereich Physik, Technische Universit\"at Kaiserslautern, 67663 Kaiserslautern, Germany}\\
    \small{3. Materials Sciences Division, Lawrence Berkeley National Laboratory,}\\
    \small{Berkeley, CA 94720, USA}\\
}
\date{}

\maketitle
\thispagestyle{plain}
{

\textbf{{A complex quantum system can be constructed by coupling simple quantum elements to one another. For example, trapped-ion  \cite{brown_coupled_2011,harlander_trapped-ion_2011} or superconducting \cite{pashkin_quantum_2003} quantum bits may be coupled by Coulomb interactions, mediated by the exchange of virtual photons.  Alternatively quantum objects can be coupled by the exchange of real photons, particularly when driven within resonators that amplify interactions with a single electromagnetic mode \cite{pellizzari_decoherence_1995}.  However, in such an open system, the capacity of a coupling channel to convey quantum information or generate entanglement may be compromised \cite{buchmann_multimode_2014}.  Here, we realize phase-coherent interactions between two spatially separated, near-ground-state mechanical oscillators within a driven optical cavity.  We observe also the noise imparted by the optical coupling, which results in correlated mechanical fluctuations of the two oscillators.  
Achieving the quantum backaction dominated regime opens the door to numerous applications of cavity optomechanics with a complex mechanical system \cite{woolley_two-mode_2013,hartmann_steady_2008,buchmann_multimode_2014,bhattacharya_multiple_2008,verlot_scheme_2009,seok_multimode_2013,stannigel_optomechanical_2012}.  Our results thereby illustrate the potential, and also the challenge, of coupling quantum objects with light.
}}

\clearpage

Cavity optomechanical systems comprised of a single mechanical oscillator interacting with a single electromagnetic cavity mode \cite{aspelmeyer_cavity_2014} serve useful quantum-mechanical functions, such as generating squeezed light \cite{brooks_non-classical_2012,purdy_strong_2013,safavi-naeini_squeezed_2013}, detecting forces with quantum-limited sensitivity \cite{schreppler_optically_2014} or through backaction-evading measurement \cite{suh_mechanically_2014}, and both entangling and amplifying mechanical and optical modes \cite{palomaki_entangling_2013}.  Systems containing several mechanical elements offer additional capabilities. In the quantum regime, these systems may enable two-mode backaction-evading measurements \cite{woolley_two-mode_2013}, creation of non-classical states \cite{hartmann_steady_2008}, fundamental tests of quantum mechanics \cite{buchmann_multimode_2014,bhattacharya_multiple_2008,seok_multimode_2013}, correlations at the quantum level with applications in high-sensitivity 
measurements  \cite{verlot_scheme_2009}, and quantum information science \cite{stannigel_optomechanical_2012}.  Realizing these proposed functions requires multiple-element cavity optomechanical systems in which quantum mechanical optical force fluctuations  dominate over thermal  and technical ones.

\begin{figure}[ht]
	\centering
	\includegraphics[width=\linewidth]{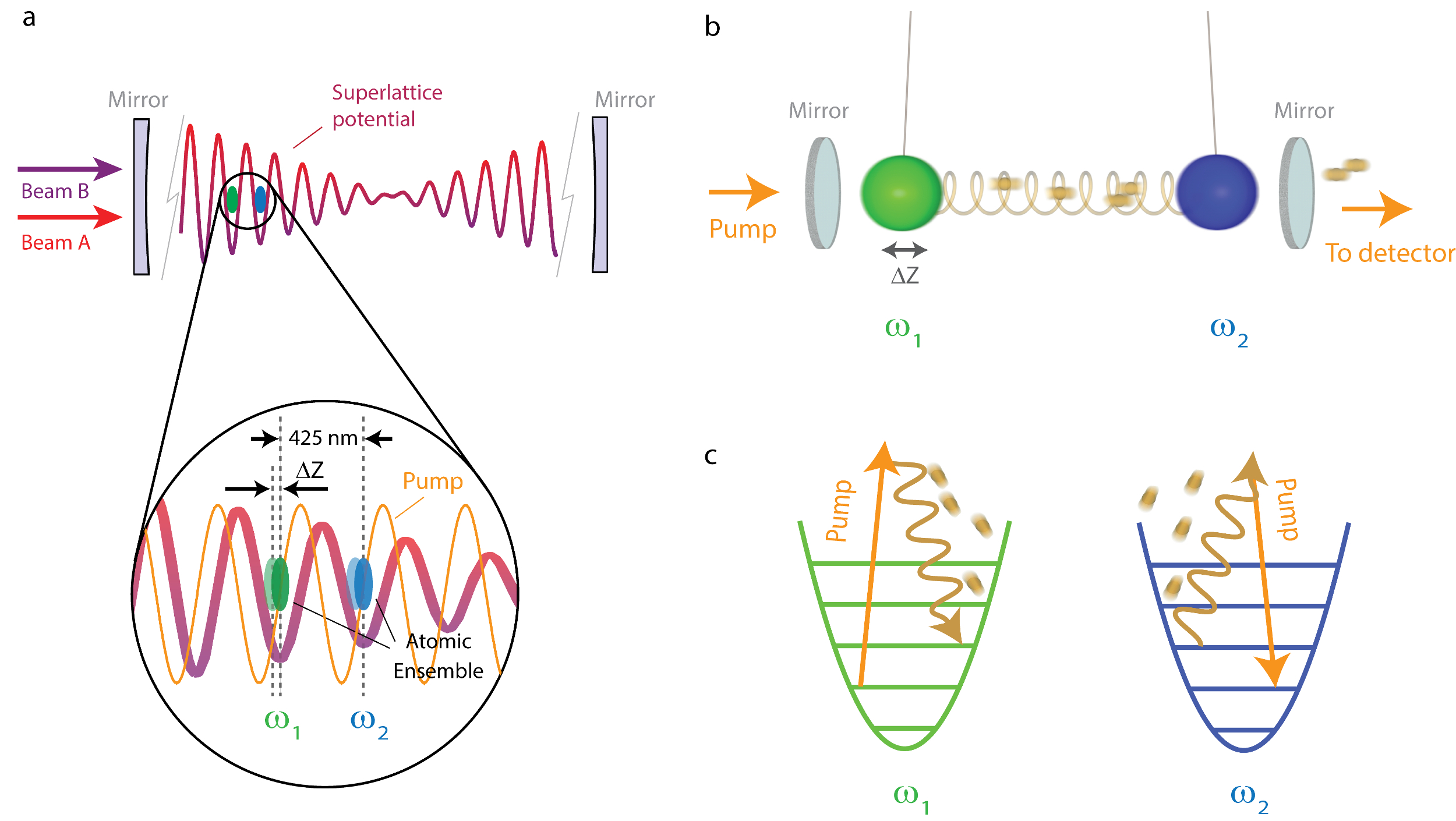}
	\dcaption{\textbf{Schematic setup.} \textbf{a} Two addressable oscillators are created by trapping ensembles of cold atoms in two sites of a superlattice potential in a cavity. The center-of-mass motion of each cloud of atoms is linearly coupled to the pump light, creating a system of two optomechanical oscillators. \textbf{b} Illustration of the realized system of coupled oscillators. Scattered pump photons induce coupling between the elements, and at the same time leave the cavity leading to backaction noise. \textbf{c} In a quantized picture, cavity photons act as the force-mediating particle for the interaction. Pump light is Stokes scattered off one element, generating a cavity photon that is absorbed through anti-Stokes scattering by the second element (and vice versa). Illustrations are not to scale.}
	\label{setup}
\end{figure}

An important new feature in these multi-mechanical systems is photon-mediated forces between mechanical elements.  Consider a driven cavity containing two mechanical elements with linear optomechanical coupling (Fig.\,\ref{setup}a).  Each element experiences radiation pressure proportional to the number of intracavity photons.  The displacement of one element changes the cavity resonance frequency, causing a change in the intracavity photon number, and thereby modifying the force on the second element.  In this manner, an effective optical spring is established between the mechanical elements (Fig.\,\ref{setup}b). A quantized picture clarifies the role of cavity photons as the force-mediating particle for this interaction: pump light is Stokes scattered off one element, generating a cavity photon that is absorbed through anti-Stokes scattering by the second element, and vice versa \cite{pellizzari_decoherence_1995} (Fig.\,\ref{setup}c).  This cavity-mediated force has been shown to cause hybridization \cite{
lin_coherent_2010,massel_multimode_2012,shkarin_optically_2014} and synchronization \cite{zhang_synchronization_2012} of, and incoherent transfer of energy between \cite{shkarin_optically_2014}, mechanical modes.  Cavity-mediated long-range interactions are also exhibited by the mechanical instability of an atomic gas within a transversely pumped optical cavity \cite{black_observation_2003,baumann_dicke_2010}.

However, in such an optically driven system, the cavity photons serve not only as force carriers but also as information carriers:  The light emitted from the cavity permits measurements on the mechanical elements inside the cavity, leading to measurement backaction noise and limiting the capacity of cavity-mediated interactions to transfer quantum states between the coupled elements \cite{buchmann_multimode_2014}.  A present challenge is first to identify and then to ameliorate or to exploit this incoherent cavity-mediated force. Thermal and technical noise prevented the observation of such incoherent backaction in previous experiments.

Here, we study a cavity optomechanical system containing two non-degenerate, spatially separated and well isolated mechanical oscillators, each comprised of a trapped ultracold gas of $^{87}$Rb atoms.  Applying cavity pump light that couples the mechanical elements, we observe light-induced forces through the redistribution of optomechanical cooperativity and, thereby, of radiation-pressure quantum noise between the different modes of motion.  Additionally, we perform time-resolved measurements of the mechanical states of the two oscillators, with sensitivity near the standard quantum limit, to reveal the phase-coherent, oscillatory exchange of energy between the elements, and also the buildup of incoherent backaction noise that accompanies the cavity-mediated coupling.

Following previous work \cite{botter_optical_2013}, we trap around 900 atoms in each of two adjacent wells of a superlattice potential, formed by two standing-wave optical fields that are resonant with TEM$_{00}$ modes of a Fabry-Perot optical resonator and far detuned from atomic resonances.  The deep superlattice potential prevents the exchange of atoms between the two ensembles.  Optomechanical interactions are introduced by driving the cavity with light detuned by -42 GHz from the D2 atomic resonance.  The driving frequency is close to another TEM$_{00}$ cavity resonance, the frequency of which varies linearly with the center-of-mass displacement of each of the trapped atomic ensembles along the cavity axis.  The linear optomechanical interaction strength is quantified by the optomechanical cooperativities of the two oscillators, given as $C_i = 4 \bar{n} g_{i}^2 / (\kappa \Gamma_i)$, where $g_{i}$ is the single-photon/single-phonon coupling frequency for the $i^{\mathrm{th}}$ oscillator, and $\kappa = 2 
\pi \times 1.82$ MHz is the cavity half-linewidth.  The transmission of this pump/probe light through the cavity is detected by a heterodyne optical receiver.  At a typical average cavity photon number $\bar{n} = 2$, $C_{1,2} \simeq 2$ allowing for single-shot detection of the motion of each oscillator near the standard quantum limit \cite{schreppler_optically_2014}.  Probing at this intensity reveals the resonance frequencies ($\{\omega_1, \omega_2\} \approx 2 \pi \times \{110, 116.4\}$ kHz) and linewidths ($\Gamma_{1,2} \approx 2 \pi \times 1.5$ kHz) of the center-of-mass oscillations of the two ensembles (Supplementary Section S1).  The superlattice potential allows us to tune the mechanical frequencies to be either non-degenerate (as here) or degenerate (see Methods).

The motion of the two ensembles becomes coupled when the pump light is tuned away from the cavity resonance.  In the unresolved sideband regime ($\kappa \gg \omega_{1,2}$), the optical force $F_i$ on oscillator $i$ varies with the oscillator positions $z_{1,2}$ as $F_i = - \sum_j k_{i,j} z_j$, where the optical spring constants $k_{i,j}$ are given by
\begin{equation}
k_{i,j} = - \frac{\hbar g_i g_j}{Z_{\mathrm{HO},i} Z_{\mathrm{HO},j}} \frac{2 \bar{n} \Delta_\mathrm{pc}}{\kappa^2 + \Delta_\mathrm{pc}^2}.
\end{equation}
Here, $\Delta_\mathrm{pc}$ is the difference between the pump light and cavity resonance frequencies, $Z_{\mathrm{HO},i} = \sqrt{\hbar / 2 M_i \omega_i}$ quantifies zero-point motion and $M_i$ is the total mass of the atoms in each lattice site (Supplementary Section S2).

\begin{figure}
	\centering
      \includegraphics[width=.6\linewidth]{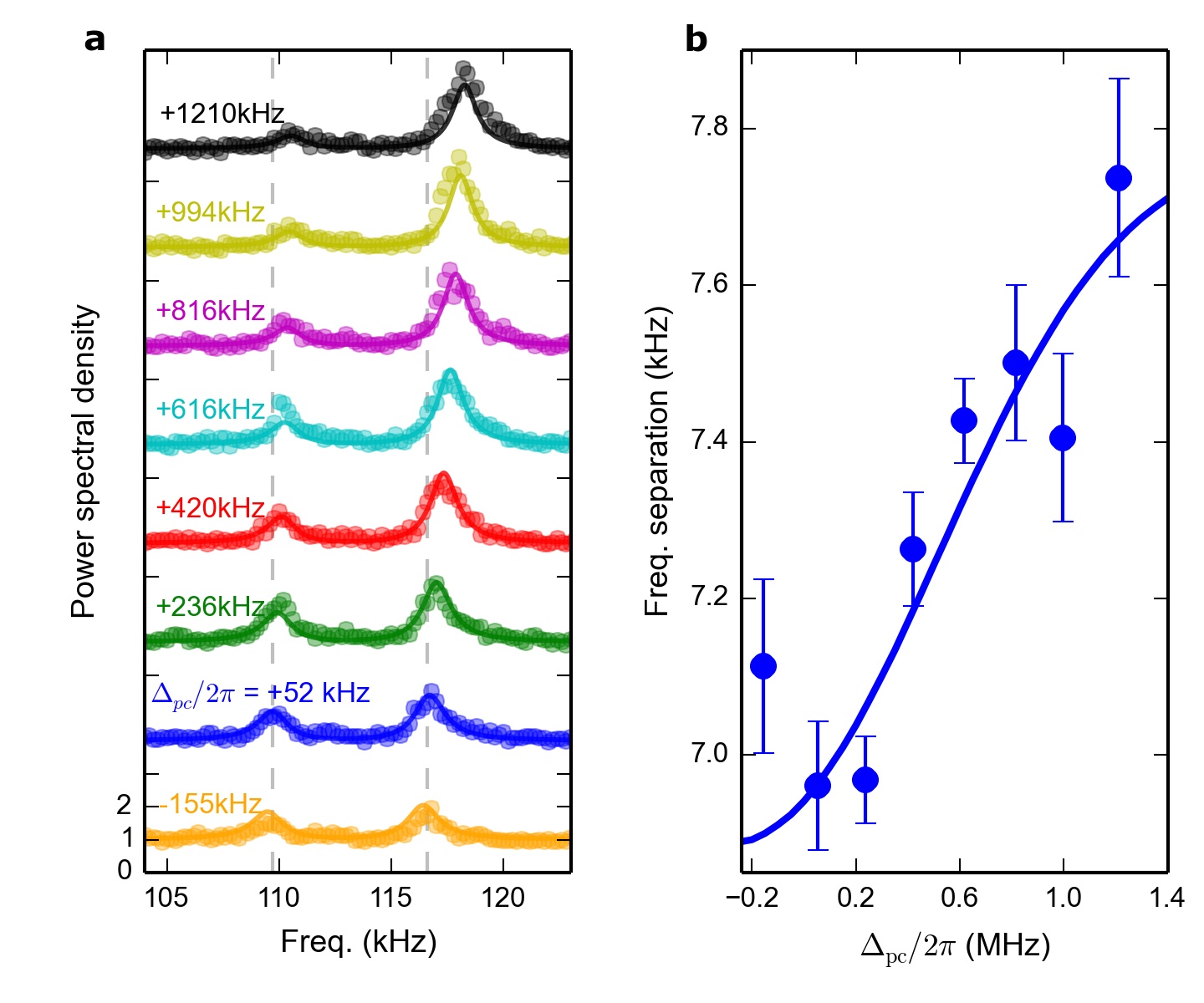}
	\caption{\textbf{Coupling of two shot noise driven oscillators.}  The cavity containing two distinct atomic ensembles is pumped with light at constant intracavity intensity ($\bar{n} \simeq 5$) and variable detuning $\Delta_\mathrm{pc}$ from cavity resonance.  \textbf{a} The symmetrized sideband power spectrum of the transmitted pump light, calculated from a 5-ms-long measurement and averaged over about 250 experimental repetitions, is shown scaled to the shot noise power seen away from mechanical resonances.  Spectra are offset by 3 shot-noise units from one another for clarity.  Optical coupling of the oscillators results in a high-cooperativity normal mode at high frequency and a low-cooperativity mode at low frequency.  \textbf{b} The coupling increases the frequency separation between the modes. Bars indicate standard fitting errors.  Solid lines show theoretical predictions.
}
\label{incoherent_coupling}
\end{figure}

Owing to these optical forces, the mechanical normal modes, describing uncoupled motion of each of the two oscillators in the absence of the optical spring, become increasingly coupled.  As the optical spring gains strength, one normal mode evolves so that its position variable matches the sum of oscillator positions weighted by optomechanical coupling strengths.  The optomechanical cooperativity of this ``bright'' normal mode becomes collectively enhanced.  The remaining ``dark'' normal mode becomes increasingly decoupled from the cavity.  The optical coupling also increases the frequency separation between the mechanical resonances.

We observe both these effects, the renormalization and frequency separation of the mechanical modes, by monitoring the symmetrized mechanical sideband spectrum of light transmitted from the cavity (Fig.\ \ref{incoherent_coupling}).  We examine the case $\Delta_\mathrm{pc} > 0$, where the optical spring leads to mechanical stiffening \cite{corbitt_all-optical_2007}.  At constant intracavity photon number ($\bar{n} \simeq 5$), we vary the pump detuning, increasing the optical spring constant up to $|k_{1,2}| \approx 1 \times 10^{-12}$ N/m. We quantify the resulting coupling strength for our system by $\Omega = \frac{|k_{1,2}|/M}{(\omega_1+\omega_2)/2} \simeq 2 \pi \times 2$ kHz, where for simplicity we approximate $\Omega$ by employing the average mass $M$ of both oscillators (Supplementary Section S3).  The steady-state spectra, with the illumination time much longer than the mechanical damping time, reveal the motion of the intracavity mechanical elements driven by zero-point fluctuations, 
thermal noise (at a level of around $\nu_\mathrm{th} = 1.5$ phonons), and radiation pressure quantum fluctuations.

At zero detuning, we observe two uncoupled mechanical resonances with comparable line strength. At non-zero detuning, the frequency difference between the two mechanical resonances increases. The increase is relatively small because the optical coupling in our experiment is rather weak ($\Omega < \omega_2 - \omega_1$); regenerative mechanical oscillations \cite{kippenberg_analysis_2005} at higher optical power prevent us from increasing $\Omega$ further at steady state. The detuned pump also leads to a difference in the power of the two mechanical sidebands, with the higher frequency resonance becoming significantly stronger.  The strength difference reflects not only that the mechanical modes are renormalized so that the cavity output is more sensitive to motion of the higher frequency normal mode, but also that this brighter mode has larger position fluctuations since it is more sensitive to radiation pressure fluctuations.  A linearized model of cavity optomechanics incorporating both these compounding 
effects
matches well to the data (see Supplementary Section S4).

The cavity-mediated force should allow for the coherent exchange of energy between the two oscillators.  To examine this exchange, we perform two-dimensional spectroscopy, such as used to study vibrational coupling in molecules and solids \cite{ernst_principles_1987,cho_two-dimensional_2009}, consisting of three steps.  First, we excite a coherent state of motion (at a level of a few $Z_\mathrm{HO} \simeq 0.8 $ nm) on just the low-frequency oscillator by resonantly modulating the intensity of one of the superlattice trapping beams. During this time, the cavity is probed with on-resonant light ($\bar{n} = 2$) so that the mechanical oscillators are not coupled.  Second, we turn on the optical coupling between the oscillators, at the level of $\Omega \approx 2 \pi \times 4$ kHz, by tuning the pump/probe light away from resonance ($\Delta_\mathrm{pc} = 2 \pi \times 1.4$ MHz) and increasing its intensity ($\bar{n} \approx 8$), holding the light at these settings for a variable time $\tau_c$.  Third, we turn off 
the optical spring and record the phase modulation of resonant cavity probe light ($\bar{n} = 2$) to characterize the mechanical state (Supplementary Section S5).

\begin{figure*}
	\centering
	\includegraphics[width=\linewidth]{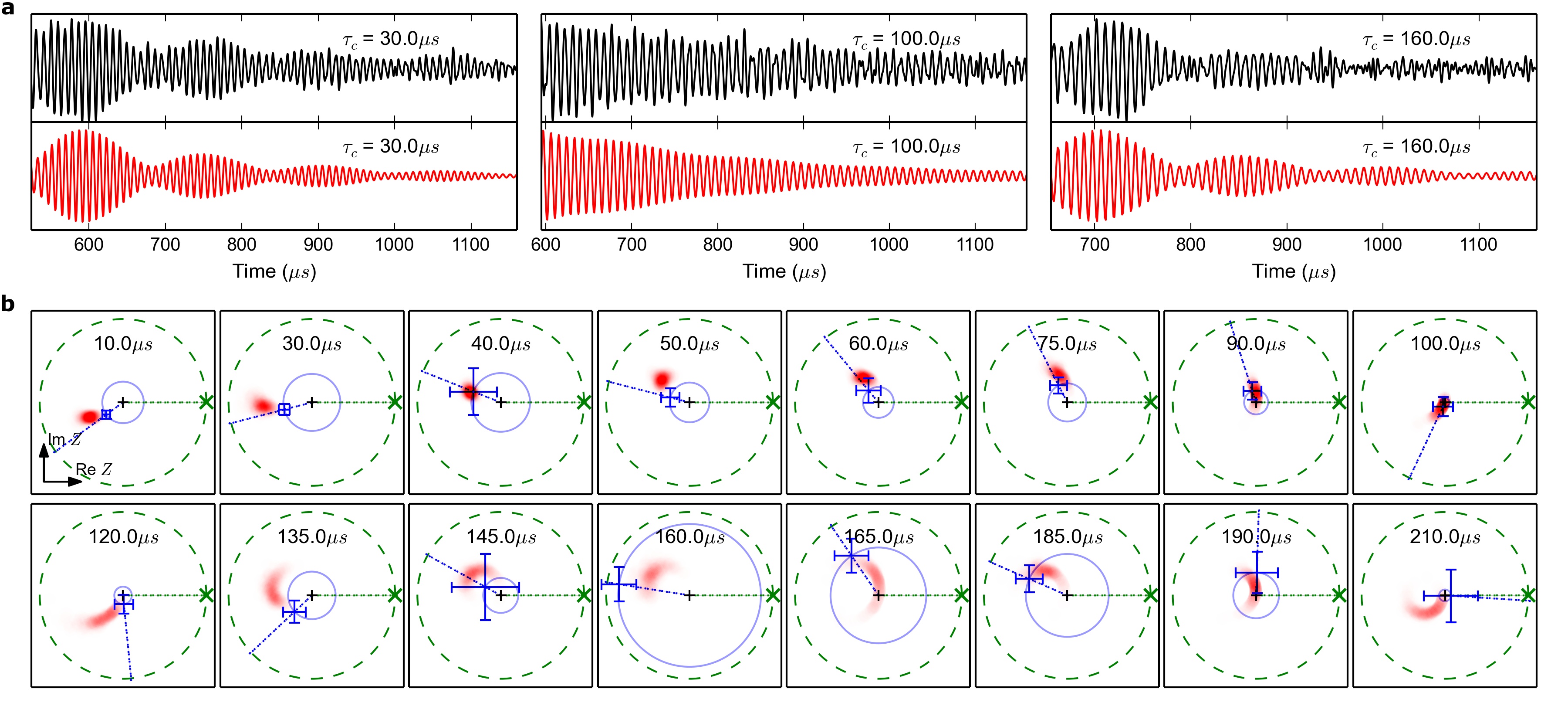}
	\caption{\textbf{Phase-coherent energy exchange.}  \textbf{a} The optical cavity probe measures a weighted sum of the displacement of the two mechanical elements.  After one element is coherently driven and an optical link between the modes is established, the beat note in the mechanical ring-down signal shows coherent oscillations of both elements.  Data (black) for each $\tau_c$, averaged over about $10^4$ experimental repetitions, show reasonable agreement with zero-free-parameter numerical simulations (red). \textbf{b} Phase-space plots indicating the average relative state of the high- (blue) and low-frequency (green) oscillators just after applying the optical coupling for varying times $\tau_c$ show two cycles of energy exchange.  Numerical simulations take into account experimental uncertainties (red shaded area) and show reasonable agreement. Bars denote standard statistical errors.}
\label{coherent_coupling}
\end{figure*}

Averaging the phase modulation signal over many experimental repetitions, we observe a time-domain beat note in the mechanical ring-down signal (Fig.\ \ref{coherent_coupling}a) showing that the coherent excitation is now shared by both oscillators.  Applying matched data filters to this time trace, as demonstrated in Ref.\ \citenum{palomaki_coherent_2013} for a one-oscillator system and detailed in the Supplementary Information (Section S6), we obtain measurements of both quadratures of motion of each mechanical element at the time when the optical spring is turned off (Fig. \ref{coherent_coupling}b).  We observe a cycle time for the exchange of energy between the oscillators of about 100 $\mu$s.  The maximal transfer of energy occurs after a half-cycle of oscillatory exchange, at the level of $A_2^2/(A_1^2+A_2^2) \approx 0.2$ where $A_i$ is the excitation amplitude of oscillator $i$.  We compare these quadrature measurements to numerical simulations (Supplementary Section S7).  The agreement with data is 
reasonable given the effects of systematic uncertainties in $\bar{n}$ and $\Delta_\mathrm{pc}$, and fluctuations in $M_i$ and $g_i$.

Our experiment demonstrates that two mechanical elements with no direct contact can be coupled to one another controllably through cavity optomechanics.  We identify two conditions required for this controlled spring force to induce fully coherent quantum evolution of the coupled mechanical system.  First, one requires the coupling rate to be larger than the thermal mechanical decoherence rate $(\nu_\mathrm{th} +1) \Gamma_\mathrm{M}$.  This condition is marginally satisfied in our work.  Second, one requires coherent evolution of the optical field acting as the force carrier.  In our system, the non-zero cavity linewidth implies that the field leaks from the cavity, providing a means to observe the state of the optical spring. The force noise resulting from this observation limits the coherence of the mechanical system.

We observe the consequence of this force noise by applying the aforementioned matched filters to each individual repetition of the three-step experimental sequence described above (Fig.\ \ref{noise}).  Measurements before the oscillators are driven and coupled characterize their initial state.   The distribution of quadrature measurements is centered about zero.  The excess variance in the quadrature measurements above the optical shot-noise level reflects the incoherent initial excitation of both oscillators at the level of 2.5 phonons.  After the oscillators are coherently excited and optically coupled, the quadrature measurement distributions are offset from the origin, reflecting the coherent motion of both elements.  Additionally, the variance of these measurements increases, showing that backaction noise during the application of the optical spring incoherently excites both oscillators.  The incoherent excitation strength is different for the two oscillators, with the high-frequency oscillator gaining 
energy faster than the low-frequency oscillator.  This difference occurs because the bright coupled mechanical mode is more strongly disturbed by backaction, and, in the weak-spring regime explored in this work, this bright mode corresponds to larger displacement of the high-frequency oscillator than of the low-frequency oscillator. Our observations are in good agreement with expectations from zero-free-parameter theory (see Supplementary Section S7).

\begin{figure*}
	\centering
	\includegraphics[width=\linewidth]{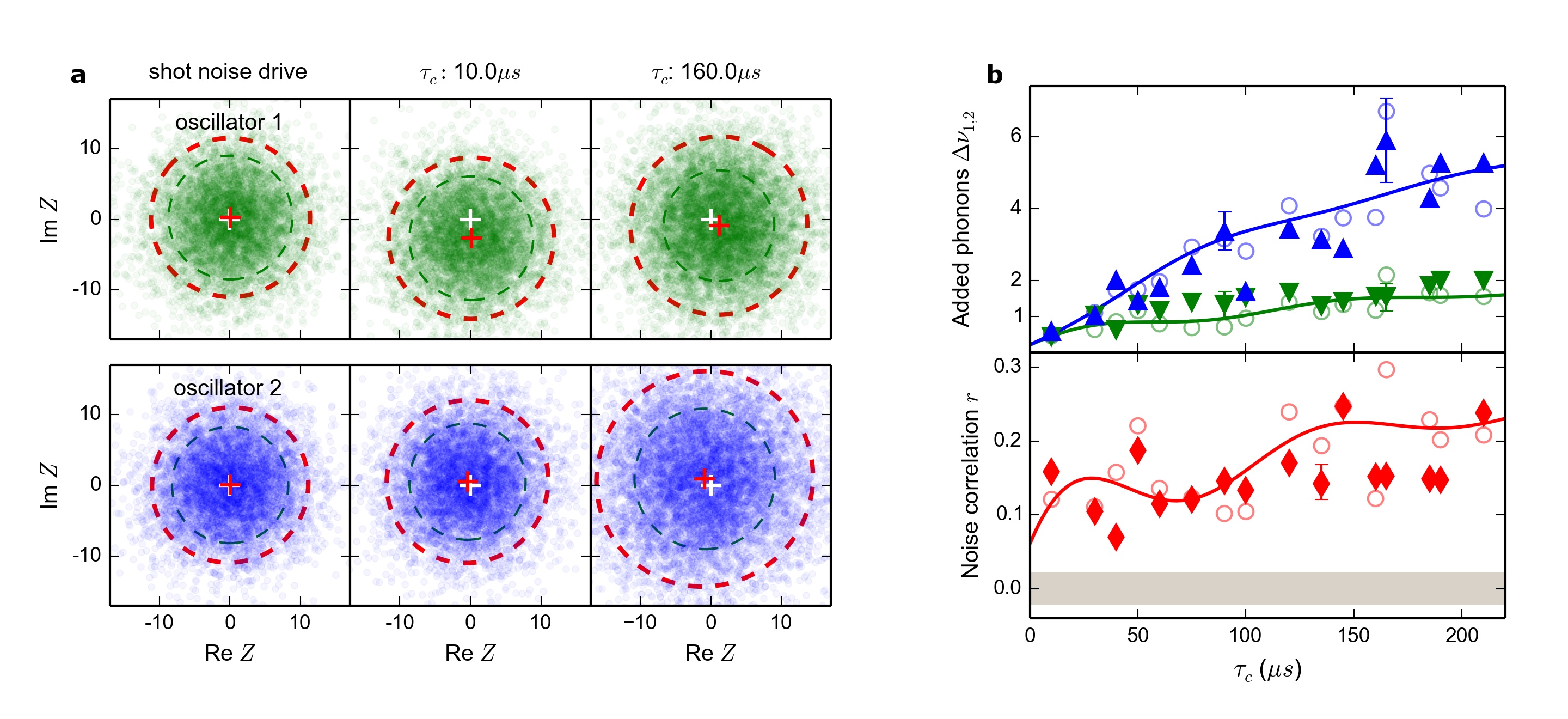}
	\caption{\textbf{Buildup of correlated backaction noise during optical coupling.}  \textbf{a} Distributions of about $ 10^4$ quadrature measurements on the two oscillators under three conditions: no coherent excitation or coupling (left) or coherent excitation and coupling of duration $\tau_c$ (center and right).  Dashed lines indicate 1-$\sigma$ variance ellipses for signal (red) and optical shot noise (black); the excess signal variance quantifies the incoherent excitation.  Optical coupling increases the incoherent excitation level, plotted in $\textbf{b}$ as the number of added phonons for the high- ({\color{blue}$\blacktriangle$}) and low-frequency ({\color{darkgreen}$\blacktriangledown$}) oscillator.  The measured data show agreement to zero-free-parameter theory calculated for each $\tau_c$ (open symbols); see Supplementary Information.  The coupling leads to correlations in the displacements of the two oscillators, quantified by the 
covariance $r_{}$ ({\color{red}$\blacklozenge$}) between their quadratures. Correlations show the expected trend, slightly reduced below zero-free-parameter expectations (open symbols). The measured 1-$\sigma$ correlations for uncoupled oscillators are indicated by grey shaded range. Solid lines indicate theory for average experimental parameters.   Bars show typical 1-$\sigma$ standard statistical errors; a plot with all error bars can be found in the Supplementary Information.}
\label{noise}
\end{figure*}

Interestingly, the backaction noise results in correlated fluctuations of the two oscillators.  These correlations, quantified by examining the covariance between motional quadratures of the two oscillators, grow in strength during the application of the optical spring. We clearly observe the build-up of correlations that is larger than the measured level of uncertainty for uncoupled oscillators, and is in reasonable agreement with zero-free-parameter theory (Fig.\ \ref{noise}).

Fluctuations of the optical spring force are found to add one phonon of incoherent excitation within about $30 \, \mu s$, a time shorter than the coherent coupling time $(\Omega/2\pi)^{-1} \simeq 250 \, \mu s$.  This comparison implies that for the conditions of our experiment -- pump light detuned about one half-linewidth from the cavity resonance -- the force linking the two mechanical oscillators cannot be used to exchange non-classical states of motion, such as phonon Fock states, between the oscillators.  However, this limitation is not fundamental, and may be mitigated by driving the cavity further from its resonance frequency \cite{buchmann_multimode_2014}, or through post-selection or feedback using light exiting the cavity. 

{
\small
\setlength{\bibsep}{2pt}

{\raggedright 
\paragraph*{Acknowlegements} This work was supported by the Air Force Office of Scientific Research and NSF.  N.S.\ was supported by a Marie Curie International Outgoing Fellowship, J.K.\ and S.S.\ by the U.S. Department of Defense through the National Defense Science and Engineering Graduate Fellowship program, and L.B. by the Swiss National Science Foundation.
\paragraph*{Author Contributions} Experimental data were taken by N.S., J.K. and S.S., L.B. developed the theoretical model. All authors were involved with experimental design, data analysis, and production of the manuscript.
}

}
}

\clearpage

\setlength\oddsidemargin{0.50in}
\setlength\topmargin{0.0in}
\setlength\textheight{8.0in}
\setlength\footskip{0.75in}
\onecolumn{
\setlength\hsize{5.5in}
\setlength\columnwidth{\hsize}
\setlength\linewidth{\hsize}
\setlength\baselineskip{16pt}

\section*{Methods}

\setcounter{equation}{0}
\setcounter{figure}{0}
\setcounter{page}{1}
\renewcommand{\thepage}{\hspace{-1.75in}\roman{page}}
\renewcommand{\bibnumfmt}[1]{[S#1]}
\newcommand{\citesup}[1]{\bibpunct{S}{}{,S}{s}{}{}\cite{#1}\bibpunct{}{}{,}{s}{}{}}

\textbf{Preparation of optomechanical oscillators.} Sample preparation is similar to the procedure described in Ref.~\citenum{botter_optical_2013}. After laser cooling and subsequent evaporative cooling in a magnetic chip trap, a sample of $^{87}$Rb atoms is magnetically transported into the cavity. The atoms are then first transferred into a (single color) optical lattice, created by exciting a TEM$_{00}$ mode of the cavity at a wavelengths of $\lambda_\mathrm{a} = 843 $ nm, with degenerate oscillation frequencies in all lattice sites. To create separately addressable oscillators, we excite a second TEM$_{00}$ mode at $\lambda_\mathrm{b} = 862 $ nm. The resulting superlattice (illustrated in Article Fig.\,1) induces an adjustable frequency splitting between neighboring sites, which we tune in this work to be around 6.4 kHz. We perform a final stage of evaporative cooling from the optical superlattice trap. The thermal phonon occupation in the superlattice (with the probe light off) is calibrated by 
employing 
time-of-flight measurements of the atomic cloud to $\nu_\mathrm{th} \approx 1.5$ phonons in the relevant axial direction of the lattice. 

The atom population of the superlattice wells is controlled by carefully positioning the magnetic trap before ramping up the optical lattice. The number of occupied wells is adjusted by the spatial extent of the cloud, tuned by rf-evaporation. For this work, we typically load each of two sites of the optical lattice with about 900 atoms. The total number of atoms $N_\mathrm{a}$ is derived from the average shift of the cavity resonance frequency $\Delta \omega_\mathrm{c} = N_\mathrm{a} g_0^2/(2 \Delta_{\mathrm{ca}})$ due to the presence of atoms, where $g_0$ is the single-atom cavity QED coupling and $\Delta_{\mathrm{ca}}$ the detuning of the probe light from atomic resonance. \\

\noindent\textbf{Optomechanical interaction.} Information about the motion of both oscillators is extracted by optomechanical interactions with the probe beam, detuned by $\Delta_{\mathrm{ca}} \simeq - 42 $ GHz to the D2 transition ($k_\mathrm{p} = 2\pi/(780$ nm)). The optomechanical interaction depends on the gradient of the probe light at the position $z$ of the atoms, quantified by $\sin(2k_\mathrm{p} z)$. Choosing the two optical lattice sites closest to the maximum probe light gradient, we create two oscillators with comparable (nearly maximal) linear optomechanical interaction strength. The non-zero curvature of the probe light causes residual quadratic coupling \citesup{purdy_tunable_2010}. We measure this additional  contribution to the trapping potential by the probe independently to be about [$\omega_\mathrm{q,1} \simeq  - 2 \pi \times 40$ ($\omega_\mathrm{q,2} \simeq  + 2 \pi \times 208$)] Hz/photon for the low (high) frequency oscillator. The fact that one oscillator experiences trapping whereas 
the other anti-trapping by the pump/probe field confirms the loading position around the point with maximum linear optomechanical coupling. \\

\section*{Supplementary Information}

\noindent\textbf{S1: A gas of cold atoms as optomechanical oscillator\\} 
With the preparation described in the Methods, we create a system where the motion of the trapped atoms is linearly coupled to the intracavity pump light. The cavity senses the collective center-of-mass mode of this motion along the cavity axis, whereas all other modes average out and are not measured. The collective center-of-mass motion of the cold gas can be thought of as equivalent to, for example, the motion of a SiN membrane and other realizations of optomechanics. This analogy establishes the description of our system as an optomechanical oscillator \citesup{purdy_tunable_2010,brahms_optical_2012}. We note that the equivalence of different realizations of optomechanical systems is reflected in a common theoretical description (see Ref. S2, for example). \\

\noindent\textbf{Extraction of oscillator properties.} Fits of the mechanical sidebands observed in the phase quadrature of the optical heterodyne detection when probed on cavity resonance are employed to extract the mechanical damping rates $\Gamma_i$, resonance frequencies $\omega_i$ and the peak height of the incoherent response $P_i$. The optomechanical cooperativity $C_i$ is quantified for each oscillator by 

\begin{equation}
 C_i = -(\nu_{\mathrm{th},i} + 1/2) + \sqrt{(\nu_{\mathrm{th},i} + 1/2)^2 + \frac{2P_i(\omega_i)/S_\mathrm{SN}-1}{4\epsilon}}
\end{equation}

\noindent where $S_\mathrm{SN}$ is the total shot-noise power spectral density and $\epsilon \approx 0.05$ is the efficiency of the heterodyne detection \cite{schreppler_optically_2014}. The single-photon/single-phonon coupling strength $g_i$ for each oscillator is then determined by $C_i = 4 \bar{n} g_{i}^2 / (\kappa \Gamma_i)$, with the mean intracavity photon number $\bar{n}$. In this work, we typically prepare samples with $g_i \approx 2 \pi \times 24$ kHz. \\

\noindent\textbf{S2: Optical forces and induced optical spring\\} The optomechanical interaction of the cavity light with the oscillators can be described by an optical dipole force acting onto oscillator $i$ which is linear in intracavity photon number $\bar{n}$:
\begin{equation}
  F_i = -\frac{\hbar g_i}{Z_{HO,i}}\bar{n},
\end{equation}
with $Z_{HO,i}$ the harmonic oscillator length. At the same time, a displacement $z_j$ of an oscillator from its equilibrium position induces a change in the cavity resonance frequency by an amount $g_jz_j/Z_{HO,j}$. In a cavity driven with a constant frequency pump, this change in resonance frequency translates to a change in intracavity photon number $\bar{n}$, which we approximate for small cavity resonance frequency changes by
\begin{equation}
 \delta\bar{n} \simeq -\bar{n} \frac{2\Delta_\mathrm{pc}}{\kappa^2 + \Delta_\mathrm{pc}^2}\frac{g_j}{Z_{\mathrm{HO},j}} z_j.
\end{equation}
With this, the force on oscillator $i$ is given by
\begin{equation}
F_i =  \sum_j\frac{\hbar g_j g_i}{Z_{\mathrm{HO},i}Z_{\mathrm{HO},j}}\frac{2\bar{n}\Delta_{\mathrm{pc}}}{\kappa^2 + \Delta_{\mathrm{pc}}^2} z_j = - k_{i,j}z_j,
\end{equation}
where we have dropped constant terms and introduced the spring constant of Article Eqn.\,1:
\begin{equation}
 k_{i,j} = - \frac{\hbar g_i g_j}{Z_{\mathrm{HO},i} Z_{\mathrm{HO},j}} \frac{2 \bar{n} \Delta_\mathrm{pc}}{\kappa^2 + \Delta_\mathrm{pc}^2}. \nonumber
\end{equation} \\

\noindent\textbf{S3: Simplified classical model\\} While the individually measured parameters are used to analyse the coupled oscillator system for each experimental data set corresponding to a specific  $\tau_c$, it is illustrative to consider a simplified system which also helps to quantify the coupling strength $\Omega$. For equal spring constants $k = k_{i,j}$, but different natural frequencies $\omega_i$ and masses $M_i$, the classical equations of motion for our system of two coupled oscillators are given by

\begin{equation}
-\left(\begin{array}{cc}
M_1 & 0\\
0  & M_2\\
\end{array}\right) 
      \begin{pmatrix}
	\ddot{z}_1\\
	\ddot{z}_2
	\end{pmatrix} = 
\left(\begin{array}{cc}
M_1 (\bar {\omega} + \delta/2)^2 + k & k\\
k  & M_2 (\bar {\omega} - \delta/2)^2 + k\\
\end{array}\right)\begin{pmatrix}
	z_1\\
	z_2
	\end{pmatrix},
\end{equation}

\noindent with the mean resonance frequency $\bar{\omega} = (\omega_1+\omega_2)/2$ and the natural frequency splitting $\delta = \omega_2 - \omega_1$. For equal masses $M=M_{1,2}$, this further simplifies to

\begin{equation}
\label{eq_simp}
      -\begin{pmatrix}
	\ddot{z}_1\\
	\ddot{z}_2
	\end{pmatrix} = 
\left(\begin{array}{cc}
(\bar {\omega} + \delta/2)^2 + k/M & k/M\\
k/M  & (\bar {\omega} - \delta/2)^2 + k/M\\
\end{array}\right)\begin{pmatrix}
	z_1\\
	z_2
	\end{pmatrix}.
\end{equation}

\noindent Normal mode frequencies of the coupled oscillators are given by the eigenvalues of the matrix in Supplementary Eqn.\,(\ref{eq_simp}):

\begin{equation}
 \omega_\pm^2 = \delta^2/4 + k/M + \bar{\omega}^2 \pm \sqrt{(k/M)^2 + \delta^2\bar{\omega}^2}.
 \label{eigenvalues}
\end{equation}

The system is in the weak (strong) coupling regime, if the splitting due to coupling is small (large) compared to $|\delta|$. Corresponding to Supplementary Eqn.\,(\ref{eigenvalues}), we quantify the strength of the coupling by $\Omega = \frac{k/M}{\bar{\omega}}$. In this work, we typically operate at $\Omega \approx 2 \pi \times (2-4) $ kHz $< \delta = 2 \pi \times 6.4$ kHz, i.e. in the weak coupling regime.

The average oscillator properties for the entire data set used in this work differ by less than 15 $\%$, which justifies the use of the simplified coupling strength $\Omega$ in the discussion of general features of our system. Calculations performed to predict the experimental observations take variations of $g_i$, $M_i$ and $\Gamma_i$ into account (see below). \\

\noindent\textbf{S4: Linear amplifier model extension to two oscillators\\} An optomechanical system can be modelled as a linear amplifier. For a single oscillator, this treatment was discussed in detail in Ref. S2. In order to model the steady state spectra for two oscillators optically coupled to the same cavity field (as presented in Article Fig.\,2), we extend this model by allowing two oscillators to interact with the cavity field. In the framework of this treatment, the evolution of the field operators in frequency space can be expressed by 

\begin{equation}
       \begin{pmatrix}
	\hat{\tilde{a}}_+\\
	\hat{\tilde{a}}_-
	\end{pmatrix} = 
	\bf{F_a}\textnormal{}\left[
	\sum_{j=1}^N T_j
	\begin{pmatrix}
	\hat{\tilde{z}}_j\\
	\hat{\tilde{p}}_j
	\end{pmatrix} + 
	\sqrt{2 \kappa}
	\begin{pmatrix}
	\hat{\tilde{\alpha}}_{in+}\\
	\hat{\tilde{\alpha}}_{in-}
	\end{pmatrix} \right]
\end{equation}

\begin{equation}
       \begin{pmatrix}
	\hat{\tilde{z}}_j\\
	\hat{\tilde{p}}_j
	\end{pmatrix} = 
	\bf{F_{b,j}}\textnormal{}\left[
	T_j
	\begin{pmatrix}
	\hat{\tilde{a}}_+\\
	\hat{\tilde{a}}_-
	\end{pmatrix} + 
	\sqrt{\Gamma_j}
	\begin{pmatrix}
	\hat{\tilde{\eta}}_{in+,j}\\
	\hat{\tilde{\eta}}_{in-,j}
	\end{pmatrix} \right]
\end{equation}

\noindent where $\bf{F_a}$ is the response matrix for the optical cavity, $\bf{F_{b,j}}$ the matrix corresponding to the mechanical response of oscillator $j$ and $\bf{T_j}$ the matrix capturing the coupling of oscillator $j$ to the light field of the cavity. Photonic and phononic inputs (including noise) are symbolized by $\hat{\tilde{\alpha}}_{in+},\hat{\tilde{\alpha}}_{in-}$ and $\hat{\tilde{\eta}}_{in+,j},\hat{\tilde{\eta}}_{in-,j}$, respectively. \\

\noindent\textbf{S5: Coherent excitation and optical coupling\\} 
\noindent\textbf{Resonant coherent drive.} We study the coherent drive of oscillators at $\Delta_\mathrm{pc} \approx 0, \bar{n} \approx 2$ and therefore $k_{12} = 0$, i.e. without coupling.

First, we intensity modulate one of the superlattice beams with a rectangular pulse of only 3 cycles duration at $\omega_{\mathrm{excite}} = 2 \pi \times 110$ kHz. 
Because of the corresponding spectral width of about $37$ kHz, such a pulse is resonant with both oscillators. In Supplementary Fig.\,\ref{ringdown} the resulting ringdown signal is presented. The observed signal shows the beat note of the two oscillators. For the time domain plots, the data are filtered by a second order butterworth bandpass filter to remove technical noise and shotnoise,  (forward and backward, to preserve phase) with a bandwidth of 40 kHz, centered around the oscillators' resonance frequencies. 

Addressing only a single oscillator in a short time requires tight control over the resonance frequencies. The intensities of both lattice beams are detected after transmission through the cavity and actively stabilized. However, drifts in the detection efficiency cause the center frequencies of both oscillators to slowly vary over time. We therefore directly stabilize the low frequency oscillator to $\omega_{1} = 2 \pi \times 110$ kHz, at $\Delta_\mathrm{pc} \approx 0, \bar{n} \approx 2$. For this, we extract the deviation of $\omega_1$ from a sideband fit averaged over a few experimental repetitions. A digital proportional-integral controller feeds back to the intensity of one of the superlattice beams, thereby stabilizing the center frequency of the low frequency oscillator (bandwidth on the order of a few minutes). The frequency drift of the second oscillator is observed to be limited to about $\pm $ 200 Hz, since we only apply feedback to one  beam of the superlattice trap. With the typical mechanical 
damping 
rate of $\Gamma_{1,2} \approx 2 \pi \times 1.5$ kHz, both oscillators are stabilized to within a fraction of their linewidth, allowing for stable spectral addressing. The oscillators' frequencies are locked in this way for all experiments presented in this paper.

For single oscillator addressing, we resonantly excite the low frequency oscillator with a pulse at $\omega_\mathrm{excite} = 2 \pi \times 110$ kHz and a duration of 50 cycles. We shape the excitation pulse with a truncated Blackman envelope to suppress spectral crosstalk \citesup{kasevich_laser_1992}. The small spectral width of such a pulse ensures the residual excitation of the high frequency oscillator with $\omega_2 \approx 2 \pi \times 116.4$ kHz to be on the order of $10^{-2}$. Supplementary Fig.\,\ref{ringdown} shows the resulting ringdown signal, dominated by a single oscillator. \\

\noindent\textbf{Coupling pulses.} To turn the optical spring on and off, the probe/pump settings are varied gradually over 20 $\mu$s to prevent spurious excitation of the mechanical elements from the increase in $\bar{n}$. 

To facilitate data taking, we repeat excitation and coupling pulses in each single run of the experiment. After applying a pulse, the mechanical system is allowed to re-equilibrate under constant illumination (at $\bar{n} \simeq 2$ and $\Delta_\mathrm{pc}\simeq0$)  before the spectroscopy sequence is repeated after 2 ms. After 5 repetitions, the atoms are released from their trap and a new atomic gas is cooled and placed within the cavity. 

Choosing a sufficiently slow repetition rate allows to read out different parts of the timetrace: The first time-window covers the ringdown following the excitation and coupling. The second time-window is shifted by several ringdown times, so that any coherent excitation has damped out and each oscillator is exclusively driven by optical shot noise and in thermal equilibrium with the phonon bath. This time window provides a record of the intrinsic properties (i.e. without coupling/drive) of each oscillator. The temperature of both oscillators in this timewindow corresponds to the initial incoherent excitation. We measure this initial phonon occupation for each $\tau_c$ value and each oscillator by sideband asymmetry \citesup{brahms_optical_2012} to be typically $2.5$ phonons. With an optomechanical cooperativity of $C_{1,2} \approx 2$, corresponding to about one phonon due to backaction \citesup{brahms_optical_2012}, and a thermal base occupation of about 1.5 phonons (see above), we find good agreement with 
expectations. \\

\noindent\textbf{Coherent energy exchange rate.} The effective frequency splitting between the two oscillators, determining the coherent energy exchange rate, is increased by two effects during the coupling. First, the coupling itself increases the natural frequency splitting of $\delta/2 \pi \simeq 6.4$ kHz  by an amount of about $1.5$ kHz. Additionally, the trapping potential of the pump light further adds about $1.5$  kHz to the frequency difference due to the higher photon number during coupling (see Methods). We therefore expect a cycle time for coherent exchange of energy of around $(9.4$ kHz$)^{-1}$. Our observations are in reasonable agreement with this prediction (Article Fig.\,3). \\

\noindent\textbf{S6: Matched filter\\} We employ a matched filter to extract quantitative results from timedomain signals. First, we extract $\Gamma_{1,2}$ and $\omega_{1,2}$ from shotnoise-only driven oscillators for each $\tau_{c}$ data set. We define two template functions $e_j(t)=e^{i\omega_jt-\Gamma_jt}$. The optimal filter coefficients $\alpha_j$ are defined as the best choice to express a signal timetrace $s(t)$ in terms of the template functions, i.e. 
$$
s(t)=\sum_j\alpha_je_j(t).
$$
The inner product between a single timetrace $s(t)$ and the template functions measures their correlation and is given by
$$
\langle e_k(t), s(t)\rangle=\int_{t_{couple}}^{t_{end}} e_k(t)s(t) dt
$$
where $t_{end}$ is the end of the ringdown timewindow. Plugging the expansion of $s(t)$ into this expression gives 
$$
\langle e_k(t), s(t)\rangle=\sum_jE_{kj}\alpha_j,
$$
with $E_{kj}=\langle e_k(t),e_j(t)\rangle$. The optimal filter coefficients can now easily be extracted by the multiplication of the vector $\langle e_k(t), s(t)\rangle$ with the inverse of the matrix $E_{kj}$. This procedure removes the effects of the template functions not being orthogonal. The $\alpha_j$ are exactly the coefficients which minimize the accumulated deviation of the timetrace $s(t)$ from the expansion $\sum_j\alpha_je_j(t)$ and constitute the best estimate of the state of each oscillator in phase-space at $\tau_{c}$. A direct comparison between the distribution of filter coefficients and timetraces is given in Supplementary Fig.\,\ref{ringdown}, where we compare a time domain fit with amplitudes and phases as free parameters to the corresponding distributions of filter coefficients. The resulting phases and amplitudes agree with the matched filter results to within $10^{-3}$. Additionally, the filter coeofficients contain information about the noise of each oscillator through the phase-space 
area of its distribution.

We further scale both real and imaginary parts of the filter output to harmonic oscillators units by multiplying with $1/\sqrt{\epsilon S_{SN} C_i \Gamma_i}$ (Ref. \citenum{schreppler_optically_2014}). This procedure corrects for differences in $\Gamma_i$ and $C_{i}$ of the two oscillators and allows us to identify the filter coefficients with dimensionless harmonic oscillator displacements $Z_i$. \\

\begin{figure}
	\centering
	\includegraphics[width=\linewidth]{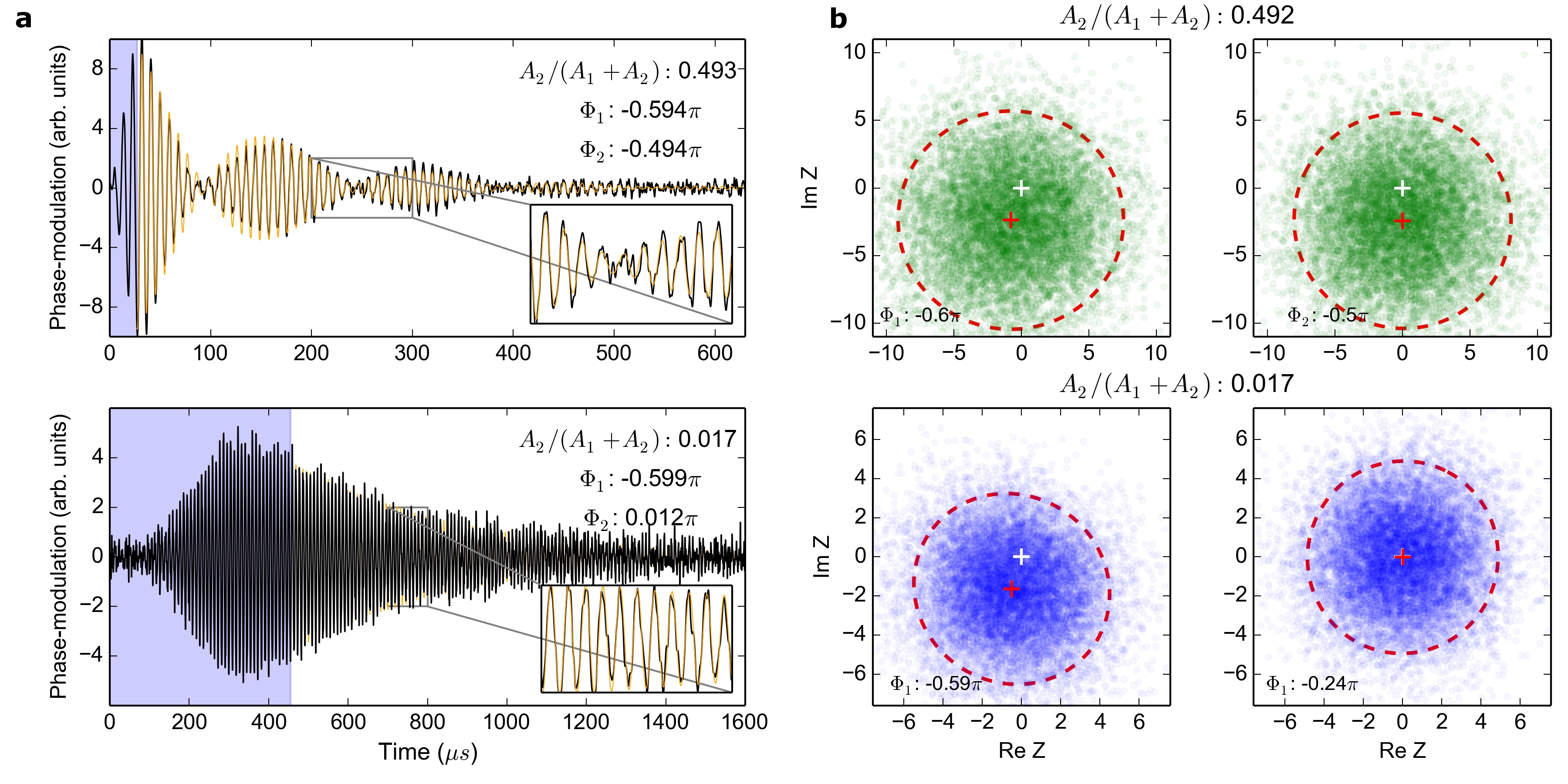}
	\dcaption{\textbf{Resonant coherent excitation without coupling, and comparison of time domain fits and matched filter.} \textbf{a} Time domain data (black) after exciting both oscillators (upper panel) and a single oscillator (lower panel). Blue shaded region indicates the timewindow in which the coherent drive is applied. Orange solid line shows the time domain fit in the ringdown timewindow. \textbf{b} Phase-space plots for the same data generated with the matched filter, in arbitrary units, after exciting both oscillators (upper panel) and a single oscillator (lower panel). Dashed red line shows $1\sigma$-error ellipse generated from the data. Extracted parameter for each approach are given. Each data set contains about $10^4$ pulses.}
	\label{ringdown}
\end{figure}

\noindent\textbf{S7: Numerical simulations for pulsed experiments\\}\label{theory} Using numerical integration, we calculate the expectation for the coherent energy exchange and the backaction noise due to coupling.

The lab-frame master equation for the density operator $\rho$ describing the cavity field and the mechanical modes including optical pump and dissipation is given by
\begin{equation}
\frac{d\rho}{d t}=\frac{-i}{\hbar}[\mathcal{H}_m+\mathcal{H}_o+\mathcal{H}_{OM},\rho]+\kappa\mathcal{L}(\hat{a})\rho+\sum_{j=1}^2\left[\Gamma_j(\nu_{\mathrm{th},j}+1)\mathcal{L}(\hat{b}_j)\rho+\Gamma_j\nu_{\mathrm{th},j}\mathcal{L}(\hat{b}_j^\dag)\rho\right],
\label{masterequation}
\end{equation}
with 
\begin{align}
\mathcal{H}_m&=\sum_{j=1}^2\hbar\omega_j\hat{b}_j^\dag\hat{b}_j\\
\mathcal{H}_o&=\hbar\omega_c\hat{a}^\dag\hat{a}+\eta\hat{a}^\dag e^{-i\omega_L t}+\eta^*\hat{a} e^{i\omega_L t}\\
\mathcal{H}_{OM}&=\sum_{j=1}^2\hbar g_{j}\hat{a}^\dag\hat{a}(\hat{b}_j+\hat{b}_j^\dag)\\
\mathcal{L}(\hat{O})\rho&=\hat{O}\rho\hat{O}^\dag-\frac{1}{2}\left(\hat{O}^\dag\hat{O}\rho+\rho\hat{O}^\dag\hat{O}\right),
\end{align}
where we have introduced the cavity resonance frequency $\omega_c$, the optical pumping rate $\eta$ and frequency $\omega_L$, the mechanical (optical) dissipation rates $\Gamma_j$($\kappa$) as well as mechanical reservoir occupation numbers $\nu_{\mathrm{th},j}$.
\par
The coupling of each oscillator to its independent, Markovian bath is captured by the third term of the master equation (\ref{masterequation}). This term leads to dissipation and decoherence of oscillator $j$ with a rate $\Gamma_j(\nu_{\mathrm{th},j}+1)$ towards a thermal state with $\nu_{\mathrm{th},j}$ phonons. Using this general formulation avoids assuming a particular dissipation or de-phasing mechanism or a preferred quadrature of motion. We chose this model because there are likely a number of microscopic relaxation and dephasing mechanisms and their details are not known. The good agreement of the experimentally obtained ringdown traces with the theoretical predictions supports our choice.
\par
Phonon annihilation and creation operators are related to the displacement operator by
\begin{equation}
\hat{z}_i=\sqrt{\frac{M_i\omega_i}{2\hbar}}\left(\hat{b}_i + \hat{b}_i^\dag\right),
\end{equation}
and the dimensionless, complex phasor $Z$ used in the main paper is identified with $\frac{1}{\sqrt{2}}\langle\hat{b}\rangle$ for each oscillator. Moving to a co-rotating frame for the mechanical modes, $\hat{b}_i\to e^{i\omega_it}\hat{b}_i$, linearizing the interaction and eliminating the cavity field yields the master equation for the reduced density operator $\rho_\mathrm{m}$. Introducing $\hat{\textbf{b}}^\top=(\hat{b}_1,\hat{b}_2,\hat{b}_1^\dag,\hat{b}_2^\dag)$ it reads
\begin{equation}
\frac{d\rho_\mathrm{m}}{dt}=\hat{\textbf{b}}^\top\mathrm{M}_1\hat{\textbf{b}}\rho_\mathrm{m}+\rho_\mathrm{m}\hat{\textbf{b}}^\top\mathrm{M}_2\hat{\textbf{b}}+\hat{\textbf{b}}^\top\mathrm{M}_3\rho_\mathrm{m}\hat{\textbf{b}},
\end{equation}
where $\mathrm{M}_i$ are the $4\times 4$ matrices
\begin{align}
\mathrm{M}_i=\left(\begin{array}{c c}
D_i^{-} & O_i^{-}\\
O_i^{+} & D_i^{+}
\end{array}\right)
\end{align}
with
\begin{align}
\left[D_1^\pm\right]_{mn}=&-g_{m}g_{n}\frac{\bar{n}e^{i\pm(\omega_m+\omega_n)t}}{\kappa+i(-\Delta_\mathrm{pc}\pm\omega_n)}\\
\left[O_1^\pm\right]_{mn}=&-g_{m}g_{n}\frac{\bar{n}e^{i\pm(\omega_m-\omega_n)t}}{\kappa+i(-\Delta_\mathrm{pc}\mp\omega_n)}\\
\left[D_2^\pm\right]_{mn}=&-g_{m}g_{n}\frac{\bar{n}e^{i\pm(\omega_m+\omega_n)t}}{\kappa-i(-\Delta_\mathrm{pc}\mp\omega_m)}\\
\left[O_2^\pm\right]_{mn}=&-g_{m}g_{n}\frac{\bar{n}e^{i\pm(\omega_m-\omega_n)t}}{\kappa-i(-\Delta_\mathrm{pc}\mp\omega_m)}\\
\left[D_3^\pm\right]_{mn}=&g_{m}g_{n}\bar{n}e^{\pm i(\omega_m+\omega_n)t}
\left(\frac{1}{\kappa+i(-\Delta_\mathrm{pc}\pm\omega_m)}+\frac{1}{\kappa-i(-\Delta_\mathrm{pc}\mp\omega_n)}\right)\\
\left[O_3^\pm\right]_{mn}=&g_{m}g_{n}\bar{n}e^{\pm i(\omega_m-\omega_n)t}
\left(\frac{1}{\kappa+i(-\Delta_\mathrm{pc}\pm\omega_m)}+\frac{1}{\kappa-i(-\Delta_\mathrm{pc}\pm\omega_n)}\right).
\end{align}
For our system, we can safely discard terms oscillating at $\pm(\omega_1+\omega_2)$ and separate the remaining terms into unitary and non-unitary parts to arrive at
\begin{align}
\frac{d\rho_\mathrm{m}}{dt}=&-\frac{i}{\hbar}\left[\mathcal{H},\rho_\mathrm{m}\right]
+\sum_{i,j=1}^4h_{ij}\left({\bf \hat{b}}_i\rho{\bf \hat{b}}_j^\dag-\frac{1}{2}(\rho{\bf \hat{b}}_j^\dag{\bf \hat{b}}_i+{\bf \hat{b}}_j^\dag{\bf \hat{b}}_j\rho)\right),
\label{mastereq}
\end{align}
with the effective Hamiltonian
\begin{align}
\mathcal{H}=\delta\Omega_1\hat{b}_1^\dag\hat{b}_1+\delta\Omega_2\hat{b}_2^\dag\hat{b}_2+J(\hat{b}_1\hat{b}_2^\dag e^{-i(\omega_1-\omega_2)t}+\hat{b}_1^\dag\hat{b}_2e^{i(\omega_1-\omega_2)t}),
\end{align}
where the frequency shifts and coupling strength are given by 
\begin{align}
\delta\Omega_i=&\mathrm{Im}\left([O_1^+]_{ii}-[O_2^-]_{ii}\right)\\
J=&\mathrm{Im}\left([O_1^+]_{12}-[O_2^-]_{21}\right).
\end{align}\\

\noindent\textbf{Simulation of motion in the time domain.} We model the coherent motion of the coupled oscillators by the classical equations of motion, taking into account the specific parameters $ \{g_j,\bar{n},\omega_j,\Delta_\mathrm{pc},\nu_{\mathrm{th},j},\Gamma_j, M_j \}$ for each $\tau_c$ data set and for each oscillator determined as described above. We further take into account the additional frequency shift due to residual quadratic coupling, see above. To account for experimental imperfections in the created coupling pulses (slightly varying transients), we feed the averaged, measured waveforms for $\bar{n}$ and $\Delta_\mathrm{pc}$ for each $\tau_c$ data set into the simulation. We obtain the time domain evolution of motion of both oscillators, as shown in Article Fig.\,3, by numerical integration. 

To estimate the uncertainty of the predicted evolution, we vary the experimental parameters according to their uncertainty by randomly choosing values from a corresponding Gaussian distribution. Repeating the simulation for a few hundred sets of random samples gives a phase-space area indicating the uncertainty, as illustrated in Article Fig.\,3 as red shaded areas. 

We note that shot-to-shot fluctuations of the loading position and therefore of the atom number in each oscillator would decrease the coupling strength and therefore the observed coherent energy transfer. However, the single shot resolution of our experiment does not allow a quantitative estimation of this effect, and the agreement with expectation suggests this effect to be small. \\

\noindent\textbf{Model for dissipation.} To calculate expectations for the backaction noise, we integrate the master equation using independently measured oscillator parameters  for each holding time $\tau_c$. 
The matrix $h$ describes the non-unitary evolution of the mechanical degrees of freedom. It results from their coupling to independent heat-baths and the additional correlated noise as a consequence of the coupling mediated by the dissipative cavity field. 
Its explicit form is 
\begin{align}
h=\left(\begin{array}{cccc}
\Gamma_-^{(1)}+\Gamma_1(\nu_{\mathrm{th},1}+1)&\Gamma_-&0&0\\
\Gamma_-&\Gamma_-^{(2)}+\Gamma_2(\nu_{\mathrm{th},2}+1)&0&0\\
0&0&\Gamma_+^{(1)}+\Gamma_1\nu_{\mathrm{th},1}&\Gamma_+\\
0&0&\Gamma_+&\Gamma_+^{(2)}+\Gamma_2\nu_{\mathrm{th},2}\\
\end{array}\right),
\end{align}
and the parameters read in the unresolved sideband regime
\begin{align}
\Gamma_\pm^{(j)}&=\frac{2g_{j}^2\bar{n}\kappa}{\kappa^2+(-\Delta_\mathrm{pc}\pm\omega_j)^2}\\
\Gamma_\pm&=\frac{\Gamma_\pm^{(1)}+\Gamma_\pm^{(2)}}{2}.
\end{align}
The lines in Article Fig.\,4 are results of numerically integrating the master equation Supplementary Eqn.\,(\ref{mastereq}) for the second order expectation values employing the experimental parameters averaged over all $\tau_c$ and including the ramps to establish and turn off the optical spring. Open symbols in Article Fig.\,4 are calculated by integrating the Supplementary Eqn.\,(\ref{mastereq}) numerically for independently measured experimental parameters $\{g_j,\bar{n},\omega_j,\Delta_{pc},\nu_{\mathrm{th},j},\Gamma_j\}$.  \\

\noindent\textbf{Calculation of added phonons.} The phase-space distribution measured by the matched filters is employed to determine the number of added phonons. Error ellipses give the phase-space areas $A_\mathrm{ref}$ for shotnoise driven oscillators and $A_\mathrm{rd}$ after excitation and coupling, the contribution of shotnoise is accounted for by calculating the output of the filter off mechanical resonance (black error ellipses in Article Fig.\,4). The number of added phonons (Article Fig.\,4b, top panel) is then given by $\Delta \nu_{1,2}=\frac{A_\mathrm{rd}-A_\mathrm{ref}}{A_\mathrm{ref}}\frac{P_+}{P_- - P_+}$, where $P_+$ ($P_-$) is the power of the blue (red) sideband. We estimate the standard statistical errors by error propagation from uncertainties in sideband power and phase-space areas. We show typical uncertainties in Article Fig.\,4 and plot the full results in Supp. Fig.\,\ref{added_phonons}a. 

\begin{figure}
	\centering
	\includegraphics[width=\linewidth]{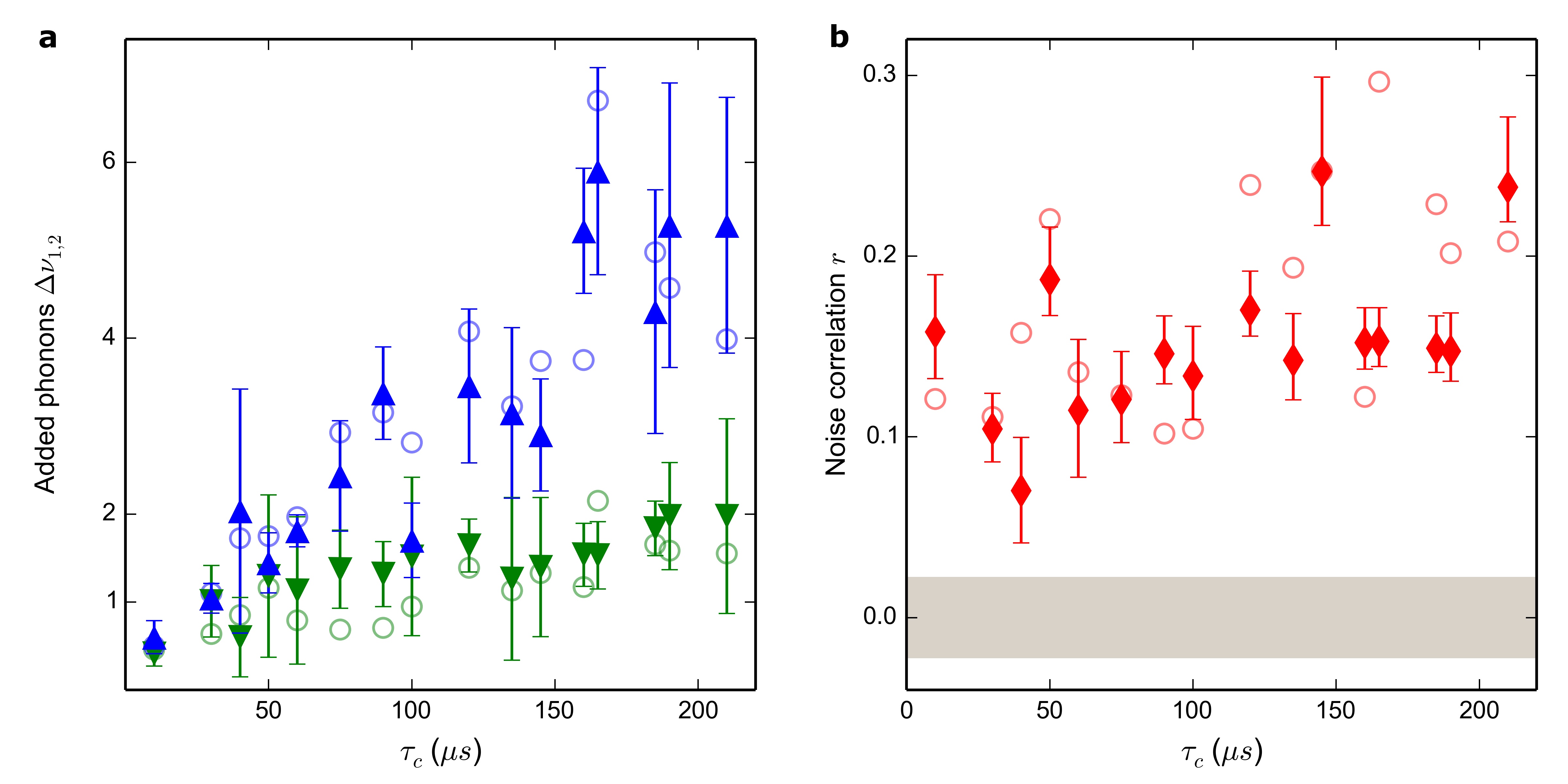}
	\dcaption{\textbf{Added phonons and correlations plotted with all errors.} \textbf{a} Article Fig.\,4b, top panel, plotted with all 1-$\sigma$ statistical errors. \textbf{b} Article Fig.\,4b, lower panel, plotted with all 1-$\sigma$ statistical errors. }
	\label{added_phonons}
\end{figure}

We test the statistical agreement between theory and experiment by calculating $\chi^2_{\mathrm{red}} = \frac{1}{N-1}\sum_N\frac{(\nu_{\mathrm{exp}}-\nu_{\mathrm{theo}})^2}{\sigma^2}= 0.97 \approx 1$, where $N$ is the number of observations. The near unity result indicates a good statistical agreement between theory and experiment. The weak oscillatory behaviour of the added phonons predicted by theory for average parameters cannot be resolved due to statistical uncertainties and to fluctuations of the experimental parameters between different $\tau_c$ runs.\\

\noindent\textbf{Noise correlations.} 
Noise correlations are quantified by
\begin{equation}
r_{\mathrm{Re}}=\frac{\langle \mbox{Re}(\hat{Z}_1) \mbox{Re}(\hat{Z}_2)\rangle}{\sqrt{\langle \mbox{Re} (\hat{Z}_1)^2\rangle\langle \mbox{Re} (\hat{Z}_2) ^2\rangle}},\label{corr}
\end{equation}
and similarly for the orthogonal quadrature  $\mathrm{Im}(\hat{Z_i})$. Theoretical predictions are calculated by integrating Supplementary Eqn.\,(\ref{mastereq}). For our system and conditions the theory predicts non-zero correlations $r_{\mathrm{Re}}=r_\mathrm{Im}\equiv r$. We can thus treat the correlations between real and imaginary parts of the coefficients as independent measurements of the correlation $r$. 
To obtain correlations from the measured filter coefficients we account for non-ideal detection efficiency and correlations introduced due to the application of matched filters. The former effect is included by subtracting the variance of filter coefficients obtained far off mechanical resonances from the variances appearing in the denominator of Supplemental Eqn.\,(\ref{corr}). The correlations introduced due to the filters are obtained by subtracting correlations of uncoupled oscillators and shot-noise in the numerator. The thus corrected correlation coefficients can be directly compared to coefficients calculated by integrating Supplemental Eqn.\,(\ref{mastereq}) for independently measured experimental parameters for different coupling times (open symbols in Article Fig.\,4b and Supp Fig.\,2b). Statistical errors are obtained by performing a Fisher transformation on the statistical variance of the sample. To verify the validity of the statistical analysis, we also extract the coefficients of the 
uncorrelated 
quantities $\langle\mbox{Re}(\hat{Z}_i)\mbox{Im}(\hat{Z}_i)\rangle$, which are scattered around zero with standard deviation given by the shaded region of Article Fig.\,4b. 

The observed correlations, averaged over all coupling times, $\bar{r}_\mathrm{exp} \approx$ 0.152 are slightly reduced from zero-free-parameter theory expectations $\bar{r}_\mathrm{theo} \approx$ 0.176. This deviation is probably caused by systematic uncertainties or shot-to-shot fluctuations leading to inhomogeneous averaging (see above). The oscillatory behaviour of the correlations predicted by theory for average parameters can not be resolved due to fluctuations of the experimental parameters between different $\tau_c$ runs.

\newpage 

}

\end{document}